# In situ X-ray area detector gain correction at an operating photon energy

James Weng,[a*] Wenqian Xu,[a*] Kamila M. Wiaderek,[a] Olaf J. Borkiewicz,[a] Jiahui Chen,[a] Robert B. Von Dreele,[a] Leighanne C. Gallington,[a] Uta Ruett [a]

[a]X-ray science division, Advanced photon source, Argonne National Laboratory, Lemont, IL 60439

**Synopsis**   A method is presented for calculating the gain map of an x-ray area detector using a series of diffraction measurements from an amorphous scatterer, allowing rapid recalibration of an area detector at any selected x-ray energy.

**Abstract**   Gain calibration of x-ray area detectors is a challenge due to the inability to generate an x-ray flat field at the selected photon energy that the beamline operates at, which has a strong influence on the detector's gain behavior. A method is presented in which a gain map is calculated without flat field measurements. Rather, a series of quick scattering measurements from an amorphous scatterer is used to calculate a gain map. The ability to rapidly obtain a gain map allows for a recalibration of an x-ray detector as needed without significant expenditure of either time or effort. Area detectors on the beamlines used, such as the Pilatus 2M CdTe or Varex XRD 4343CT, were found to have gains which drift slightly over timescales of several weeks or after exposure to high photon flux, suggesting the need to more frequently recalibrate for detector gain.

## 1. Introduction

### 1.1. Motivation

Gain calibration of x-ray area detectors at beamlines is a challenge due to the inability to generate an x-ray flat field at a selected photon energy. Gain maps are thus often collected using flat fields generated by x-ray fluorescence and then extrapolated to the photon energy used at the beamline. Detectors which have been in service for extended periods of time at beamlines often show radiation damage; pixels that are heavily exposed to high photon flux will have significantly different responses than the surrounding pixels that shows up as a burnt in pattern on the detector. Pixels which have been obscured for long periods of time, such as by a beam stop, when used again a change of the setup, have significantly different responses from used pixels and will display an artifact such as an apparent shadowing on the detector. In medical imaging using similar detectors, it is known that simply translating the detector results in erroneous apparent shadowing of an imaged object (Park & Sharp, 2015), also resulting in the need for gain corrections. Like the shadowing artifacts in medical imaging with area detectors, similar artifacts are present for diffraction measurements using an area detector; simple translation of an area detector yields significant variations of an integrated 1D pattern for powder diffraction, resulting in the observation of either false peak shifts or a detector position dependent phase change. For techniques which rely on subtle changes in baseline shifts to detect changes in structure such as pair distribution function (PDF), biological small angle scattering (bioSAXS), and contrast variation analysis, faulty measurements caused by improper detector gain calibration almost certainly leads to inaccurate later analysis. Detector gain has also, in this work, been found to drift a measurable amount on the time-scale of several weeks, potentially accumulating in such a way as to cause erroneous measurements.

In this work we present a method to rapidly measure and calculate a gain map for an x-ray area detector at the selected photon energy at which a given beamline operates. This method, along with the provided code, is not detector specific, and has no special sample or environment requirements for the measurement procedure. The only requirements for calculating a gain map in the presented work are the ability to translate the detector perpendicular to the incident beam, a calibration standard, and an amorphous scatterer. For the amorphous scatterer, common glass microscope slides are found to be sufficient for the purposes of collecting a gain map.

## 2. Calculation Approach

### 2.1. Problem Statement

The idealized 2D diffraction pattern measurement without detector errors, $S(x,y)$, can be described as the product of the measured diffraction intensity $I_m(x,y)$ and the detector gain $G(x,y)$ plus an error term $\varepsilon$ (Eq. 1).

$$S(x,y) = G(x,y) \cdot I_m(x,y) + \varepsilon \qquad (1)$$

For any diffraction experiment, $I_m$ is the measurement off the detector. $S$ is the corrected scattering pattern. The assumption is made that the detector gain, G, is constant over the time-scales relevant to a diffraction measurement; a failure of this assumption implies that the detector cannot be used for the measurements. It is also assumed that the detector response is linear; if the detector response is not linear it is likely not suitable for experimental measurements in general. For the detectors used on the beamlines at Sector 11 at the Advanced Photon Source (APS), the Perkin Elmer XRD 1621 and Varex XRD 4343CT are specified to be linear within 1% over their full scale range (PerkinElmer, 2008; Varex industrial, 2021). The Pilatus 3 X 2M CdTe has a factory software correction applied and is also specified to be linear within 1% (Dectris, 2016).

If $S$ is known, without considering the measurement error $\varepsilon$, then the gain of the detector is given by the following equation:

$$\frac{S(x,y)}{I_m(x,y)} \approx G(x,y) \qquad (2)$$

It is assumed that the errors in detector measurement, ε, are normally distributed and average to zero, so they can be ignored if enough repeat measurements are taken. In order to obtain a gain map, a signal *S* with known scattering properties must be measured. Ideally, for simplicity *S(x, y)* should be a uniform flat field, making the measured signal simply the map of pixel gain by equation 2. However, it is not necessarily possible to generate a uniform flat field at the selected x-ray energy. In this situation, a sample with known scattering properties, resulting in a signal *S* with known properties must be used. Knowledge about the sample's structure is not necessary. The generated signal *S(x,y)* must fulfill several criteria to be useful for calculation of a gain map by equation 2. It must not be sparse over the x and y; it must be primarily made up of non-zero values over x, y or G(x, y) is indeterminate at the zero values by equation 2. *S* must also be something which the detector can accurately capture; a signal with large jumps in intensity from the baseline may result in something which is not measurable within the dynamic range of the detector. The attenuation of the scattered beam by the sample has to be constant for same angles, which is true for spheres and flat objects, but not for capillaries. An amorphous scatterer producing smooth and very broad diffraction rings, such as a glass slide, fulfils these requirements.

**2.2. Description of algorithm**

For the purposes of this description, $\hat{*}$ denotes a computed estimate of $*$. A measurement is taken of an amorphous scatterer, assumed to have a radially constant scattering pattern under unpolarized incident beam, with the detector translated to several positions changing the center of the diffraction rings (figure 1). At each of these positions a calibration powder standard is measured to determine accurately center and detector to sample distance for calculating *x, y* pixel positions to *2θ* positions.

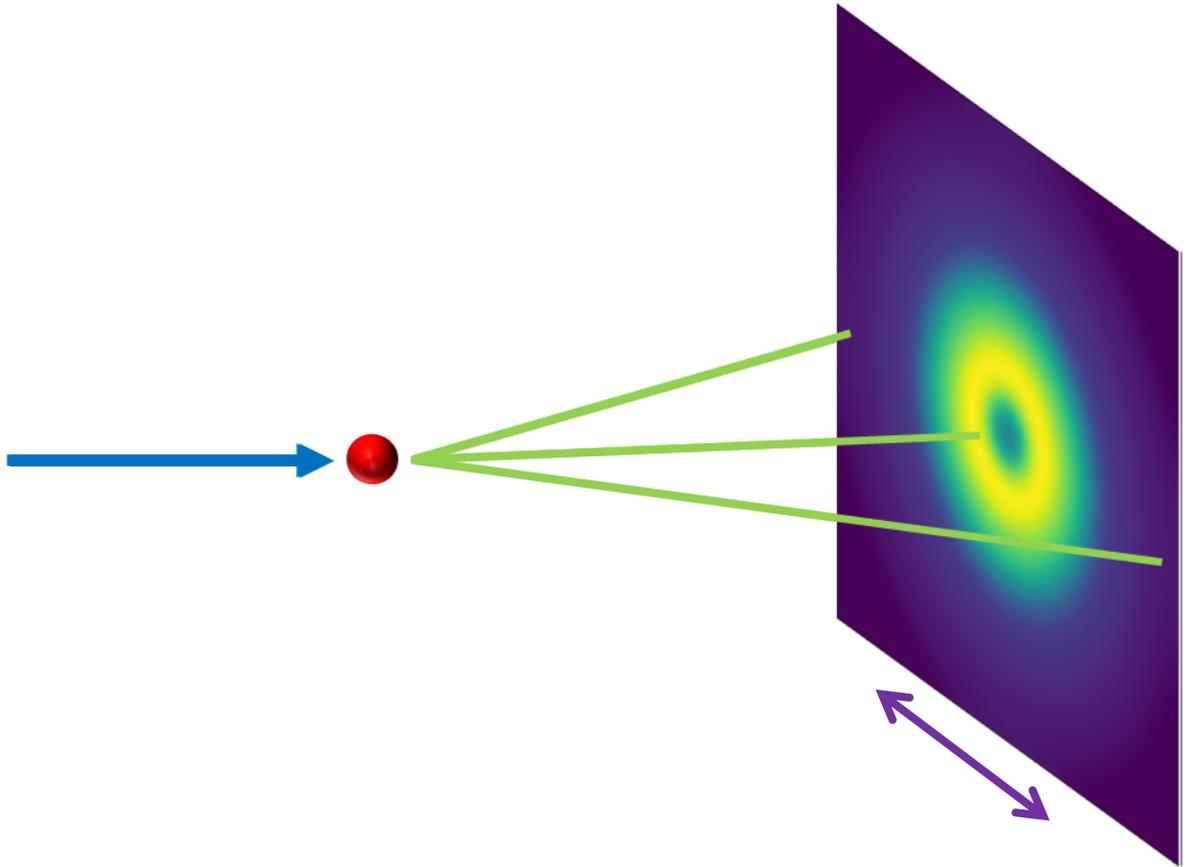

**Figure 1** An illustration of the scattering experiment used to collect data needed for a gain map calculation. Incident x-ray beam originates from the left and is shown as a blue arrow. An amorphous scatterer (red), such as glass, scatters x-rays (green) which results in a radially symmetric scattering pattern on the detector, shown on the right. The detector is translated in directions perpendicular to the incident x-ray beam (purple arrow), and multiple measurements are taken.

Using this map, from the measured signal $I_m(x, y)$ after correction of beam polarization, a radial average $I_m(r)$ of the amorphous scatterer is calculated to provide a 1D scattering pattern. A median is used as the representative radial average over the distance to the center here, as outliers are expected if individual pixels read much higher or lower than the rest of the detector prior to gain correction. The outliers are frequently seen on detectors which have sustained radiation damage. $I_m(r)$ is then mapped back to create an estimate of the signal $\hat{S}(x, y)$. From equation 2, this allows a first approximation of $G(x, y)$, $\hat{G}(x, y)$ to be calculated. This process is outlined as a flow chart in Figure 2.

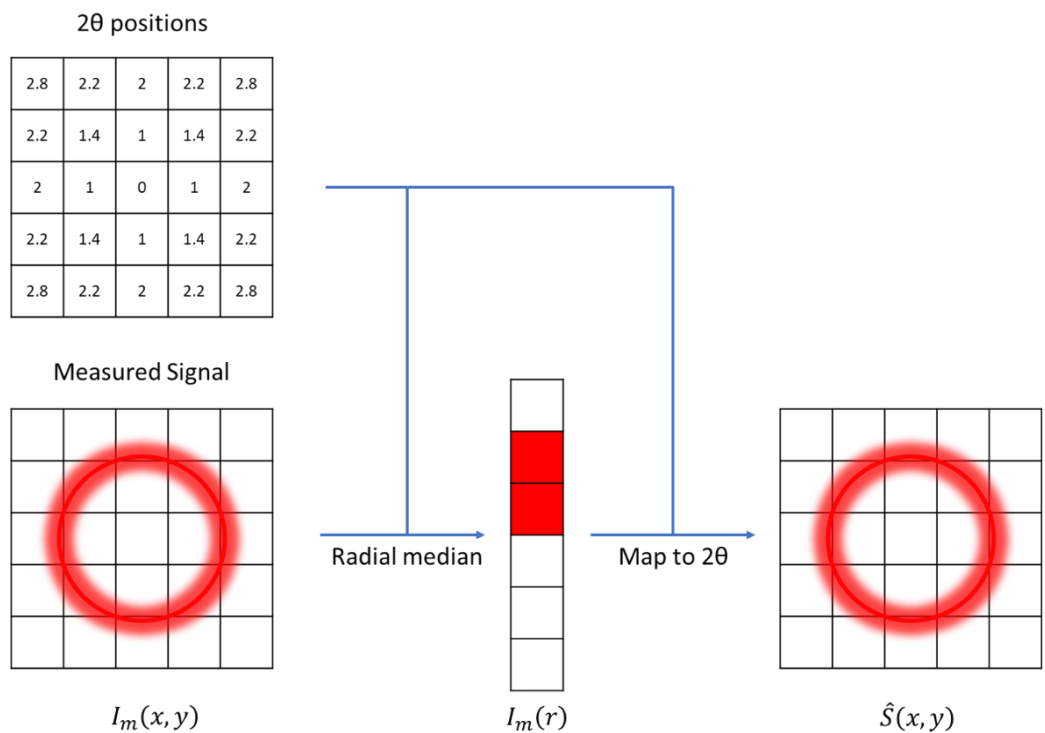

**Figure 2** Process by which the estimated idealized 2-D scattering signal is calculated by reduction of a radially symmetric 2D scattering pattern to 1D by taking a radial median and mapping it to computed 2θ positions for each pixel in the original 2D image.

The radial averages of the same amorphous scatterer with the detector translated varies slightly due to incorrect gain, resulting in an apparent slight shift in the scattering profiles in the measured 1D pattern. For example, with powder x-ray measurements the region around the usual beam stop position in the center is exposed to higher photon flux than the rest of the detector and sustains more damage over time, resulting in a larger change in gain over time at those pixels relative to the rest of the detector. For any particular experimental setup frequently measuring similar materials, different regions on the detector may consistently receive a higher photon flux. Without gain correction, it is expected that multiple radial averages with the detector translated will have shifted peaks and erroneous intensities. This is easily observed by plotting multiple radial averages of the same scatterer where the only difference is detector translation (figure 4), where peak intensity changes are enough to create apparent isosbestic points in the data that do not correspond to any real changes in the scatterer.

In order to correct for this in the final estimate of $\hat{S}(x,y)$, the amorphous scatterer is measured with the detector translated to multiple positions normal to the incident beam to measure multiple $I_m$. This procedure is outlined in figure 3.

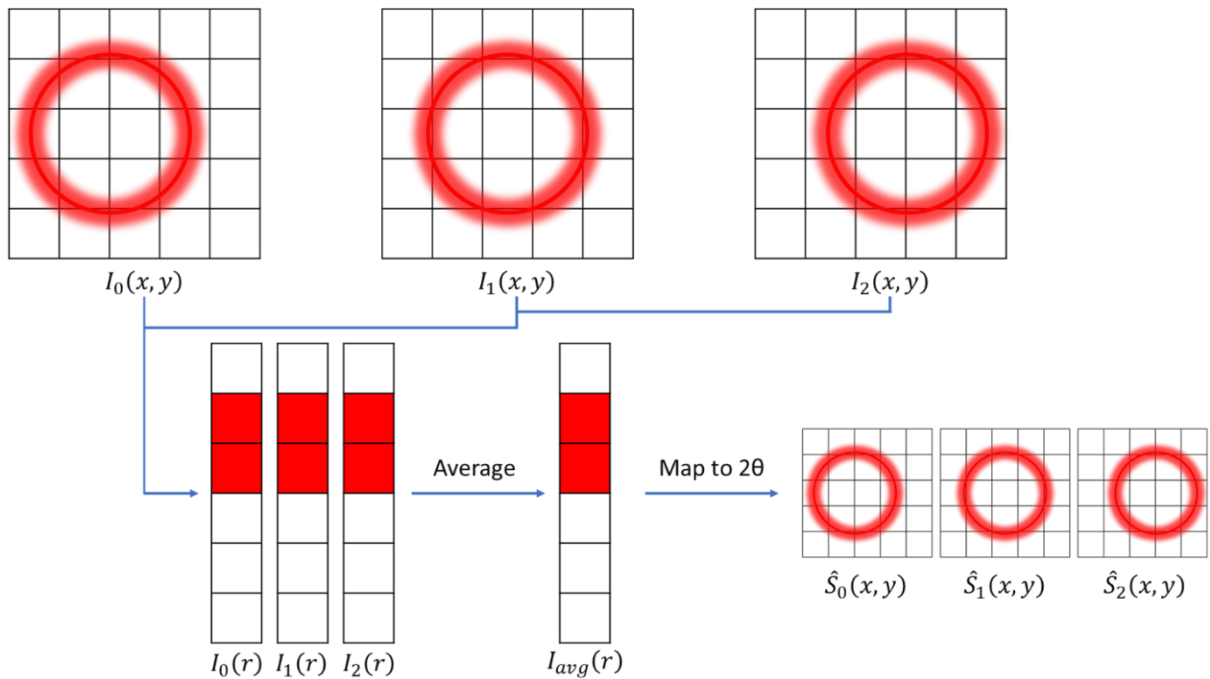

**Figure 3** Estimation of an idealized 2-D scattering signal by taking the average of multiple radial medians with the detector translated to several positions. The computation with several detector positions is needed to account for any systematic radially dependent changes in the detector gain caused by repeated measurements with the detector at the same position.

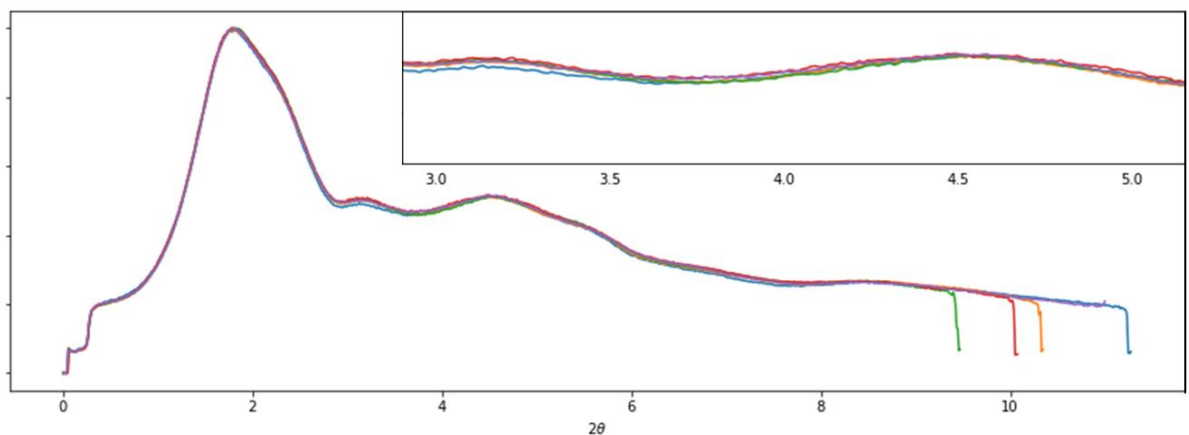

**Figure 4** Multiple radial averages of some amorphous scatterer with the detector translated slightly between each measurement showing apparent changes in relative scattering intensity when using an outdated gain map. Changes in relative intensities would lead to incorrect conclusions about the structure, particularly in systems sensitive to low spatial frequency changes in the measured pattern such as PDF measurements on liquids. Detector used was a Pilatus 2M CdTe detector.

The 1D patterns $I_m(r)$ for each translated detector position are averaged, to get a better estimate of $S$. Using the previously calculated $x, y \to 2\theta$ map, the averaged $I_{avg}(r)$ is mapped back to 2D for each measured position, providing an estimated $\hat{S}_m(x, y)$ for each scattering measurement. Using equation 2, an estimated $G_m(x, y)$ is calculated at each position $m$, which contains both null values where the detector is obscured by the beam stop, as well as a ring of erroneous values surrounding the beam stop caused by asymmetric scattering of, partial obstruction of the scattering profile by, or some other unknown artifact created by the beam stop. Both these artifacts are also visible in previous literature with similar approaches to measuring x-ray detector gain maps (Wernecke *et al.*, 2014). This process of calculating a position specific gain map $\hat{G}_m$ is shown visually in figure 5.

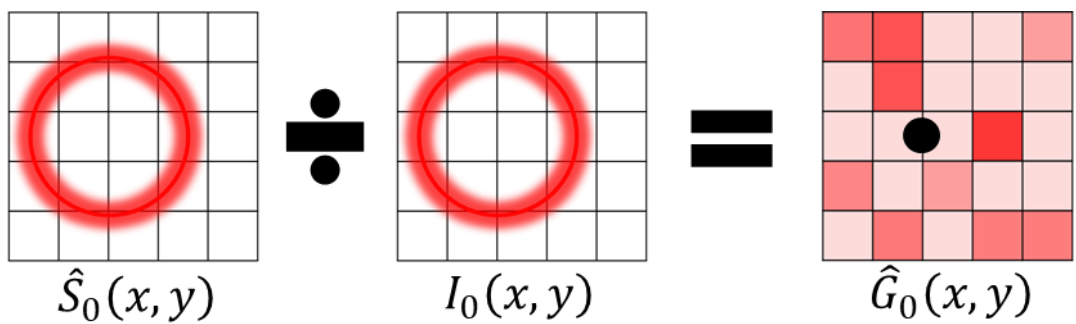

**Figure 5** Division of an idealized scattering pattern, $\hat{S}_0$, for a given position 0 by an experimentally measured scattering pattern, $I_0$, at the same position produces a position specific gain map $\hat{G}_0$. The position specific gain map has no values over any regions obscured by a beam stop, denoted by a black circle. Any asymmetric scattering in the experimental setup will appear in $\hat{G}_0$ as an erroneous apparent gain that does not correspond to an actual detector gain.

In order to remove these artifacts the median of all $\hat{G}_m(x, y)$ is computed to provide an estimated $\hat{G}(x, y)$ which contains neither null values nor scattering artifacts which originate from the asymmetrical scattering from the beamstop. The presence of such artifacts can additionally be visualized by viewing the absolute difference between a position specific gain map and the median gain map of the detector:

$$Abs[G(x, y) - G_m(x, y)] \qquad (3)$$

The absolute difference for any given $\hat{G}_m$ from $\hat{G}$, produces an image showing circular artifacts which are concentric to the beam stop position. Inclusion of the information contained at these points would lead to an erroneous computed gain map $\hat{G}$.

Multiple position specific artifacts-removed gain maps $\hat{G}_m$ are then combined to produce a gain map $\hat{G}$ which contains neither the region occluded by the beam stop nor asymmetric scattering from the beamstop. This is performed by simply taking the median of several position specific gain maps.

From here, damaged pixels are separated from the gain map. Damaged pixels are those which are identified as having apparent gains which deviate from the median gain of the detector by more than 4 standard deviations ($\sigma$). These pixels are separated to produce a gain map containing only undamaged pixels and a second map containing only damaged pixels (figure 6).

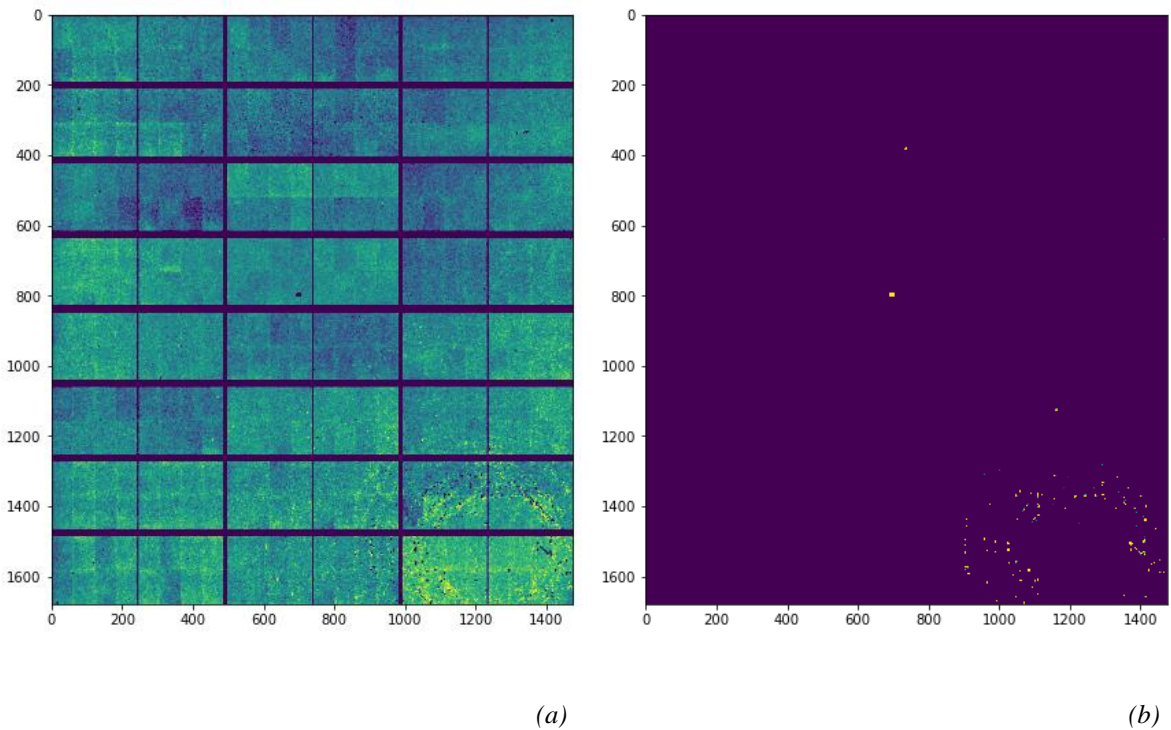

*(a)* *(b)*

**Figure 6** Calculated gain maps for a Pilatus 2M detector with the normally responding pixels (a) and heavily radiation damaged pixels (b) separated. Heavily radiation damaged pixels are defined as those with calculated gains which are 4 standard deviations away from the average gain of the detector.

The gain map of undamaged pixels then has a median filter applied in order to reduce the contribution of shot noise. The window size used for the median filter is selected to be no larger than the measured point spread of the detector. In the case of PerkinElmer 1621 or Varex 4343 detectors, the median filter window size used is 7x7 pixels, and for Pilatus 2M CdTe, the median filter window size used is 3x3 pixels. The point spread functions of the detectors are measured by illuminating a single pixel with an attenuated x-ray beam cut down to less than 1 pixel in size. For the scintillator based PerkinElmer 1621 and Varex 4343 detectors, the spread is measured to be a region of at least 10x10 pixels, while the Pilatus 2M Cd the spread is measured to be a region of at least 3x3 pixels.

A combined gain map containing gain corrections of the damaged pixels and undamaged pixels is created by summing the median filtered map of the undamaged pixels with the unmodified map of the damaged pixels. An alternative way of treating the damaged pixels is to exclude them from data analysis, and the map of damaged pixels, e.g. Figure 6b, can be used as a mask. It should be determined if the damaged pixels are providing useful measurements with some large apparent gain, or providing erroneous measurements when deciding how to handle them in later data analysis.

**2.3. Experimental**

A series of scattering patterns was collected using an amorphous scatterer with the detector placed fairly far away from the sample. For measurements taken at 17-BM-B, 11-ID-B, and 11-ID-C at the Advanced Photon source, the sample to detector distance used was 1000 mm. Detector distance was chosen such that there was still scattering intensity at the edges of the detector. The amorphous scatterer used in all experiments was a stack of 1.5mm glass microscope slides. The number of microscope slides used for each measurement was selected so that the scattering intensity across the detector was within the linear regime of the detector. Scattering patterns were taken with the detector at five different positions (SI 1). Detector translation was arbitrary and was found to not be critical for the calculation of the gain map, so long as the region occluded by the beam stop did not overlap between measurements. A calibration standard was also measured at each position so that a map of pixel to 2θ positions could be calculated using GSAS II (Toby and Von Dreele, 2013). GSAS II was also used to generate a pixel map of intensity scale due to polarization, so that the polarization induced intensity difference at different azimuth angles can be corrected. A gain map was then calculated from these measurements with the software provided at (Weng, 2022). It does not appear that the sample to detector position used is critical to calculation of a gain map, so long as the scattering intensity across the detector is not attenuated by air scattering to an intensity below the linear response of the detector. Scattering conditions used at various beamlines along with the detector used are given in Table 1. It should be noted that the detector should be allowed to sit between measurements for at least as long as the measurement in order to minimize burn in effects altering the measurement. It is also worth noting that beam polarization should be accurately determined so that polarization induced intensity difference can be correctly removed before gain map calculation. Beam polarization can be precisely obtained from 2D scattering data of an amorphous material, such as a stack of glass slides as used in this study, using a recently published method (Von Dreele and Xu, 2020).

**Table 1**

Measurement parameters for gain map calculation at various beamlines

| Beamline | Energy | Scatterer Used | Collection Time Per Position | Sample to detector distance | Detector |
| --- | --- | --- | --- | --- | --- |
| 17-BM-B | 27 keV | 4 slides | 12 minutes | 400 mm | Varex 4343 CT |
| | 51 keV | 4 slides | 12 minutes | 1000 mm | Varex 4343 CT |
| 11-ID-B | 58.6 keV | 8 slides | 3 minutes | 1000 mm | PE XRD1621[1] |
| | 86.7 keV | 8 slides | 5 minutes | 1000 mm | PE XRD1621 |
| 11-ID-C | 105.7 keV | 10 slides | 2.5 minutes | 1000 mm | PE XRD1621[1] |
| 11-ID-C | 105.7 keV | 10 slides | 2.5 minutes | 1000 mm | Pilatus 2M CdTe |

Scatter used for measurements was a stack of 1.5 mm thick glass microscope slides, 1 – PE XRD1621 detectors used were two different detectors of the same model

## 3. Results

It is observed that translation of the detector changes relative peak shapes and intensities of an amorphous scatterer without proper gain correction. Simply translating the detector, taking a measurement, and radially averaging provides 1D diffraction patterns which are not identical. Attempting to normalize these patterns based on their max intensity results in patterns with differing peak profiles. Correction of the detector data based on a measured gain map removes this problem and results in 1D patterns which are in agreement regardless of the detector position (figure 7).

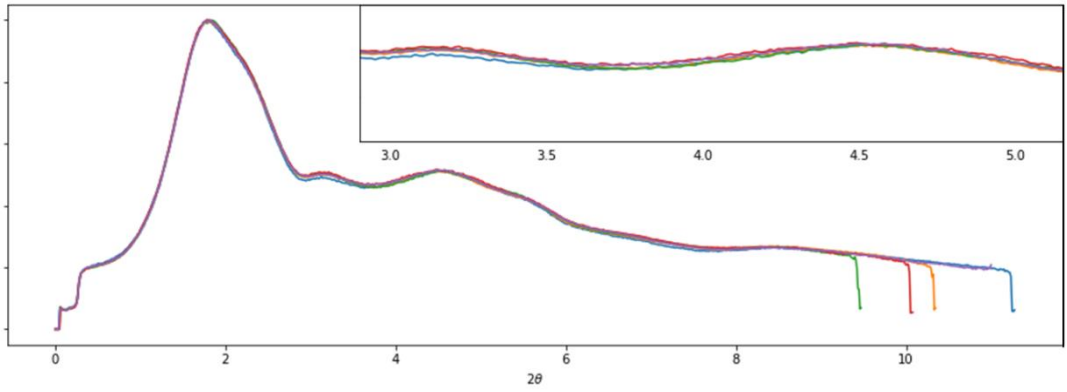

*(a)*

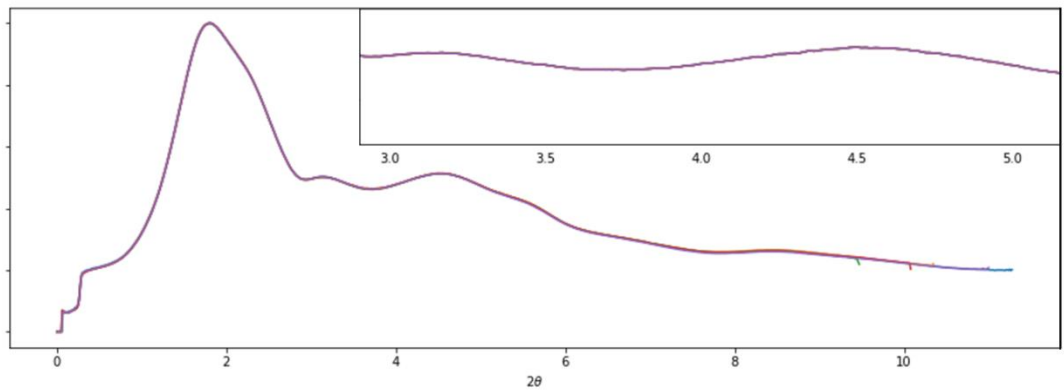

*(b)*

**Figure 7** Radial averages of an amorphous scatterer measured with a Pilatus 2M CdTe detector with the detector translated slightly between measurements that have been corrected with the factory detector gain map (a) and a recently calculated gain map (b). The measurements have different relative peak intensities when corrected with the factory gain map which cannot correspond to sample structure, indicating that the detector gain has changed over time.

For some materials, such as carbon nanoparticles, a collected diffraction pattern without gain correction results in a 1D pattern with erroneous peaks which do not correspond to a real structural feature (figure 8).

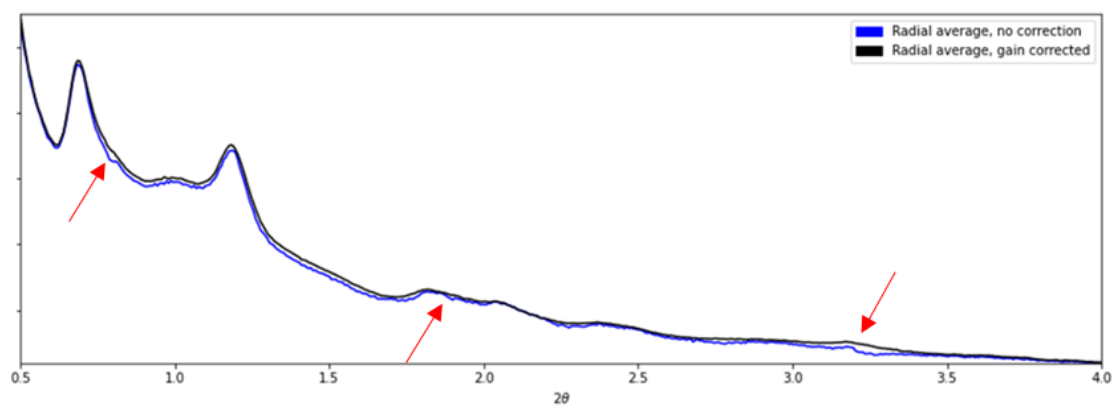

**Figure 8** Radial averages of the scattering from carbon nanoparticles without gain correction (blue) and with gain correction (black) showing the presence of small peaks that are detector gain related, rather than sample related (major peaks labelled in red). Measurements were taken on 11-ID-C at 105.7 keV with the Pilatus 2M CdTe detector.

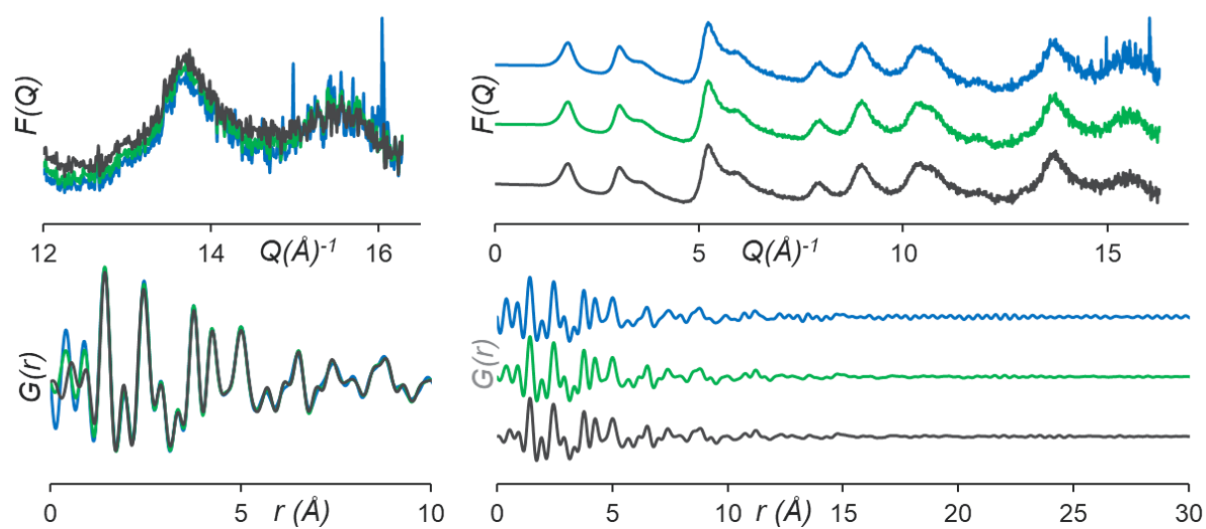

**Figure 9** F(q) functions (a) calculated from Vulcan amorphous carbon with no correction (blue), hot pixel mask (green), hot pixel mask and gain correction (black) and corresponding pair distribution functions (b)

When collecting data suitable for calculating pair distribution functions (PDF) from materials with low scattering intensity, low crystallinity samples such as amorphous carbons or liquids, proper image corrections significantly improve signal-to-noise ratio and decrease artifacts in the data. Since most amorphous samples do not contain significant features at higher Q values in the measured *I(q)*, any noise related deviation of the signal in high Q regions will be especially harmful to measurements of these samples. Figure 9a shows F(Q) data of amorphous carbon "Vulcan". This is a high surface area, low density material, which contributes less than 10% of the total scattering signal, *I(q)*, when measured in glass capillaries. In the uncorrected Vulcan measurement, the amplitude of variations across the uncorrected detector pixels is larger in magnitude than the data itself, resulting in significant noise in final PDF calculation. This creates non-physical features at low r (<1Å) regions of the *G(r)* which are almost 50% in amplitude of the highest peak. As a consequence, the fit to the uncorrected "Vulcan" sample refined against P6$_3$mc graphite structure within pdfGui software (Farrow, et al.) resulted in relatively high residuals (RW=0.41) In addition, the high frequency noise at distances higher that 10 Å overpower low amplitude features and may result in less accurate size estimations. Masking non-linearly responding pixels, eliminates most of the obvious noise lowering RW by 5%. (RW=0.36). Gain correction further reduces both high frequency and low r noise bringing down RW values to 0.28.

Detector gain maps were not found to vary significantly over the course of 12 hours of continuous measurements with the Pilatus 2M detector. However, over the course of a month the gain map of a Pilatus 2M detector was found to change enough such that the radial average of measured material was slightly different (SI 2). Corrected 2D scattering with a recent gain map shows less of a "speckle pattern" than the one corrected with an old gain map, implying that the gain of individual pixels drift slightly independent of the surrounding pixels (Figure 2b). Gain maps of the Pilatus 2M CdTe and Perkin Elmer XRD1621 detectors at 11-ID-C are shown in figure 10 and were calculated using measurements taken at 105.7 keV.

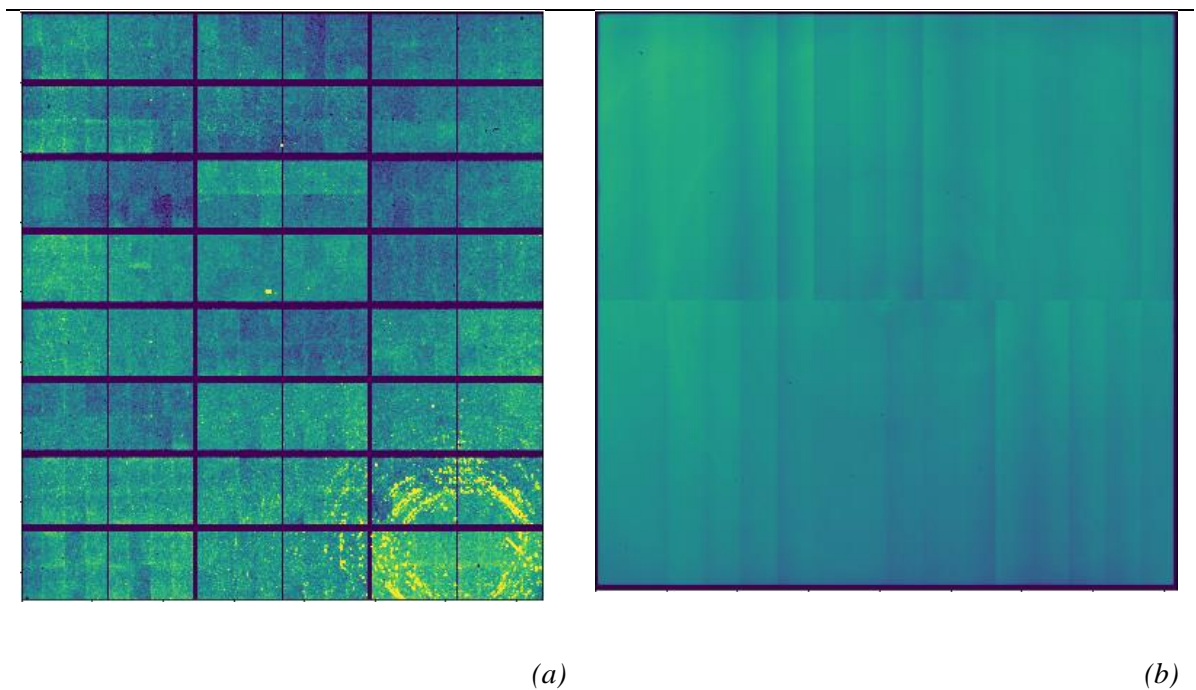

*(a)*          *(b)*

**Figure 10**  Gain maps calculated for Pilatus 2M CdTe (a) and a Perkin Elmer XRD1621 (b) detectors using data collected at beamline 11-ID-C at 105.7 keV. Energy threshold for the Pilatus 2M

CdTe detector was set to 50 keV, and gain map was found to match the map extrapolated from multiple flat field calibrations automatically calculated by the detector computer.

Displaying calculated gain maps for detectors as a height map provides a quick way to visualize regions which have been radiation damaged (Figure 11) where large spikes appear where the detector is radiation damaged. A detector with a gain map that appears as a conically shaped height map is indicative of radiation damage across the detector, and is likely a sign that the detector should be retired. The Perkin Elmer XRD1621 detector at 11-ID-B with such a gain map was additionally found to not provide accurate diffraction measurements in later tests.

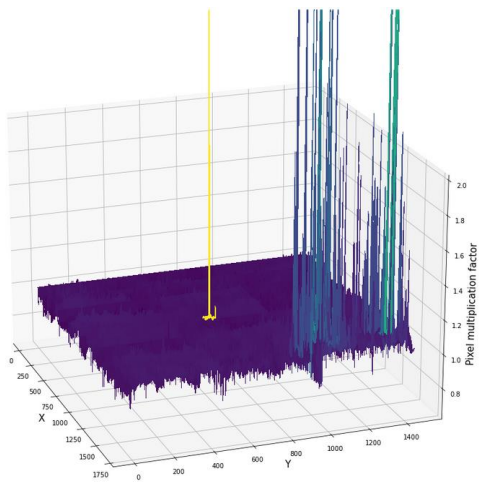

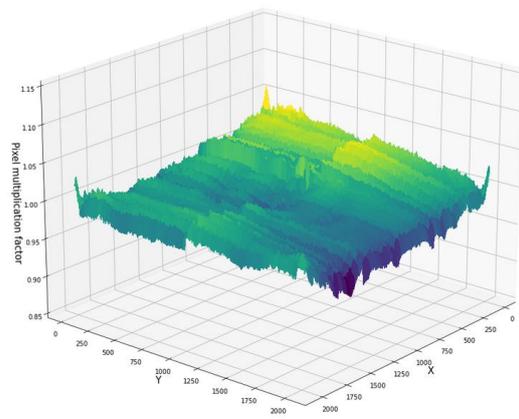

*(a)*  *(b)*

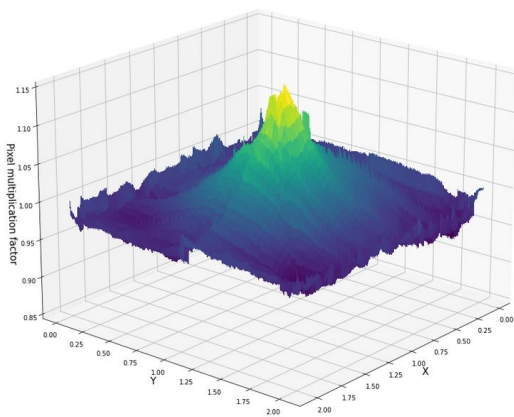

*(c)*

**Figure 11** Gain maps displayed as height maps for the Pilatus 2M CdTe detector at beamline 11-ID-C (a), Perkin Elmer XRD1621 at beamline 11-ID-C (b), and the Perkin Elmer XRD1621 at beamline 11-ID-B (c). Very large spikes in the Pilatus 2M CdTe detector are characteristic of radiation damage, with the largest spike in the center of the detector being the result of direct exposure to the x-ray beam. The large conical structure in the gain map of the Perkin Elmer XRD1621 at 11-ID-B is the result of heavy radiation damage caused by use over many years and is characteristic of a detector that should be retired. The relatively flat gain map of the Perkin Elmer XRD1621 at 11-ID-C is characteristic of a normally functioning detector.

4. Conclusions

The ability to quickly calculate a gain map using measurements easily taken at a beamline in several minutes provides a number of advantages over the usual flat field gain map calibration. The quantum efficiency of a detector varies with photon energy, resulting in a gain at each pixel specific to the x-ray energy used. By using measurements taken at the photon energy at which a beamline operates, there is no longer a need to take several flat field measurements at different photon energies using x-ray fluorescence and computing the detector gain to extrapolate to the actual photon energy used at the beamline.. Gain maps obtained by this method were found to be equivalent to those calculated from multiple x-ray fluorescence flat fields. By using measurements that are quickly obtained directly at the beamline, it becomes easier to monitor any gain changes in the detector resulting from radiation damage. Additionally, a slight drift in detector gain maps were observed over time scales of several weeks, suggesting a need to frequently recalibrate detectors for gain in order to provide accurate measurements. Experiments which rely on the measurement of subtle changes in a measured signal, such as the measurement of PDFs of liquids and amorphous materials, will benefit from a recently gain calibrated detector directly before the experiment.

**Acknowledgements** This research used resources of the Advanced Photon Source, a U.S. Department of Energy (DOE) Office of Science user facility at Argonne National Laboratory and is based on research supported by the U.S. DOE Office of Science-Basic Energy Sciences, under Contract No. DE-AC02-06CH11357.

# Supporting information

Supporting information (such as experimental data, additional figures and multimedia content) that may be of use or interest to some readers but does not form part of the article itself will be made available from the IUCr archives and appropriate databases. If possible, please include supporting material here; otherwise, separate supporting files may be uploaded upon submission of your article.

Supplementary sections, tables and figures should be numbered with a leading 'S'. The styles 'IUCr sup heading 1', 'IUCr sup table caption', 'IUCr sup figure caption', *etc*. will apply the numbering automatically.

## S1. First-level heading (style name: IUCr sup heading 1)

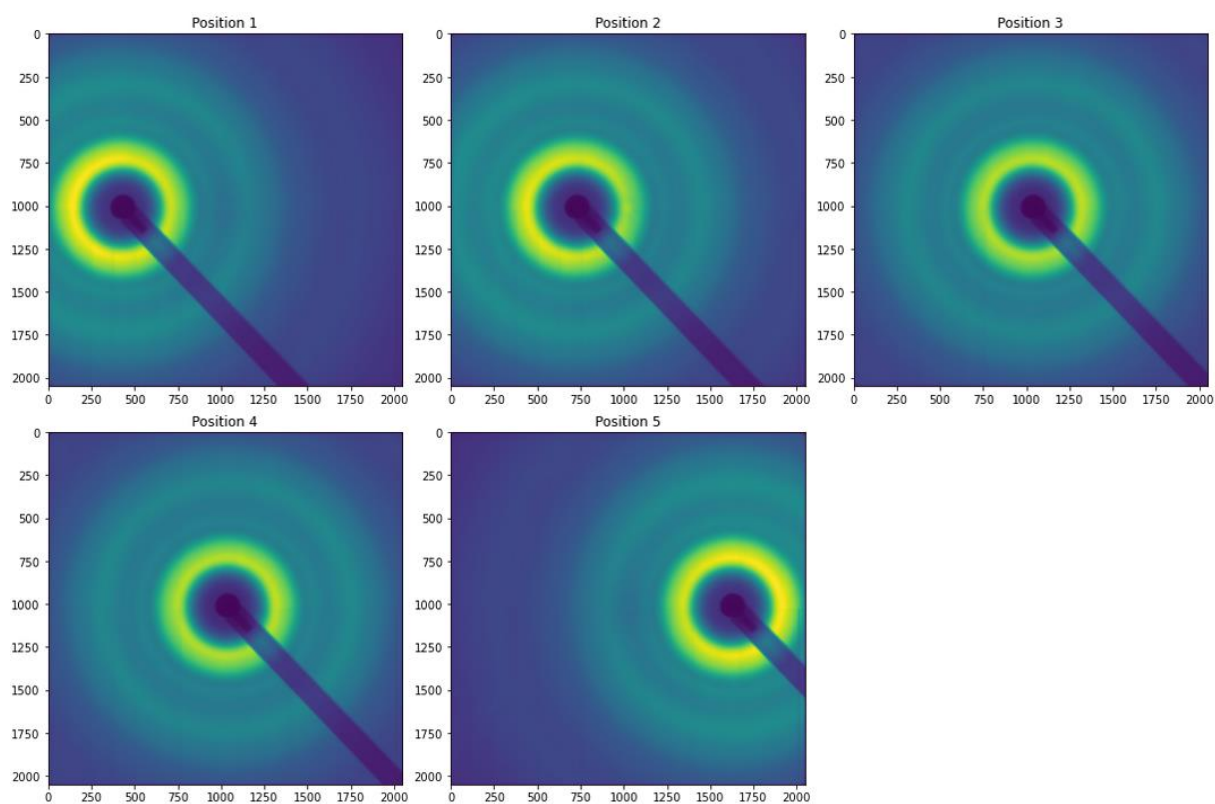

**Figure S1**  2D Diffraction measurements with an identical sample and the detector translated to five different positions used to calculate a gain map. The translation distance of the detector between the measurements is not critical, so long as the occluded beam stop regions do not overlap between the measurements.

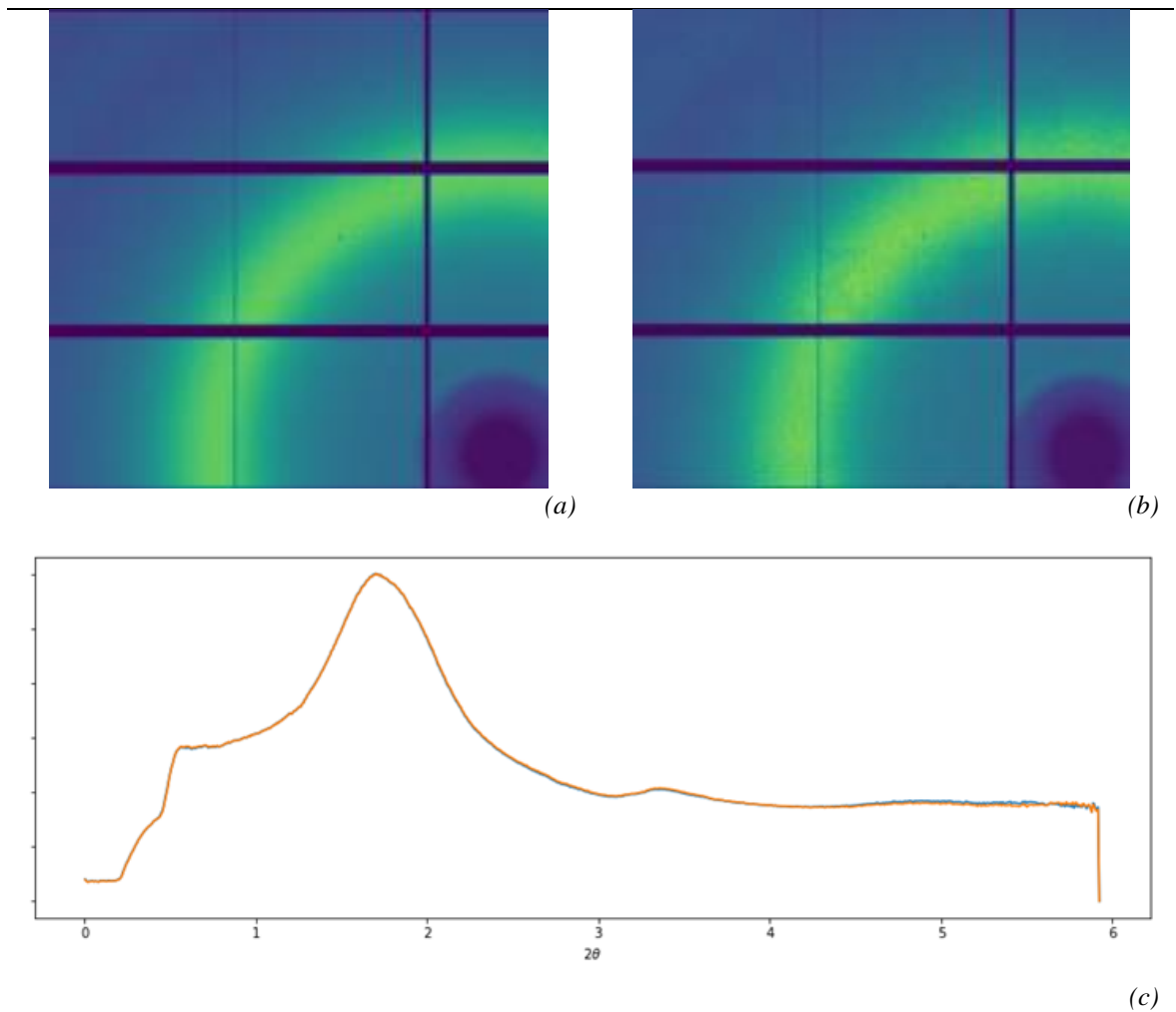

**Figure S2** Measured 2D scattering corrected using a recently collected gain map (a) and a month old gain map (b) along with the corresponding radial averages (c). Radial average from the recent gain map is shown in orange and the old gain map is shown in blue. Corrected 2D scattering with a recent gain map shows less of a "speckle pattern" than the one corrected with an old gain map, implying that the gain of individual pixels drift slightly independent of the surrounding pixels.